\documentclass{emulateapj}
\begin{document}

\title{Ultraluminous Star-Forming Galaxies and Extremely Luminous Warm Molecular Hydrogen Emission at $z=2.16$\\ in the PKS 1138-26 Radio galaxy Protocluster}

\shorttitle{Luminous H$_2$ in the PKS 1138-26 Protocluster}

\author{P. Ogle$^1$, J. E. Davies$^1$, P. N. Appleton$^2$, B. Bertincourt$^3$, N. Seymour$^4$, G. Helou $^1$}

\affil{$^1$ IPAC, California Institute of Technology, Mail Code 220-6, Pasadena, CA 91125, ogle@ipac.caltech.edu}
\affil{$^2$ NHSC, California Institute of Technology, Mail Code 220-6, Pasadena, CA 91125}
\affil{$^3$ Columbia University Astronomy Laboratory}
\affil{$^4$ CSIRO Astronomy \& Space Science, P.O. Box 76, Epping, NSW 1710, Australia}

\shortauthors{Ogle et al.}

\begin{abstract}

A deep {\it Spitzer} Infrared Spectrograph (IRS) map of the PKS 1138-26 galaxy protocluster reveals 
ultraluminous PAH emission from obscured star formation in three protocluster galaxies, including 
H$\alpha$-emitter (HAE) 229, HAE 131, and the central Spiderweb Galaxy. Star formation rates of 
$\sim 500-1100 M_\odot$ yr$^{-1}$ are estimated from the 7.7$\mu$m PAH feature. At such prodigious formation 
rates, the galaxy stellar masses will double in 0.6-1.1 Gyr. We are viewing the peak epoch of star formation for 
these protocluster galaxies. However, it appears that extinction of H$\alpha$ is much greater (up to a 
factor of 40) in the two ULIRG HAEs compared to the Spiderweb. This may be attributed to different spatial distributions
of star formation--nuclear star formation in the HAEs versus extended star formation in accreting satellite galaxies
in the Spiderweb. We find extremely luminous mid-IR rotational line emission from warm molecular hydrogen in the 
Spiderweb Galaxy, with $L$(H$_2$ 0-0 S(3))= $1.4 \times 10^{44}$ erg s$^{-1}$ ($3.7\times 10^{10} L_\odot$), $\sim 20$ 
times more luminous than any previously known H$_2$ emission galaxy (MOHEG). Depending on temperature, this corresponds 
to a very large mass of $>9\times 10^{6}-2\times 10^{9} M_\odot$ of $T>300$ K molecular gas, which may be heated by the 
PKS 1138-26 radio jet, acting to quench nuclear star formation. There is $>8$ times more warm H$_2$ at these temperatures
in the Spiderweb than what has been seen in low-redshift ($z<0.2$) radio galaxies, indicating that the Spiderweb
may have a larger reservoir of molecular gas than more evolved radio galaxies. This is the highest redshift galaxy 
yet in which warm molecular hydrogen has been directly detected. 

\end{abstract}


\section{Introduction}

Galaxy clusters are the largest virialized structures in the universe. Massive, 
X-ray selected clusters are well established at redshift $z=1$ \citep{fgh07}, and
their brightest galaxies have colors consistent with passively evolving ellipticals 
\citep[e.g.,][]{mya07}. The progenitors of such clusters must have formed at an earlier epoch, 
around the highest density peaks in the cosmos. Such protoclusters have been found around
$z = 2-5$ radio galaxies \citep{pkr00,vrm07,hdg11,khv11, tdk11}, which appear to be construction 
zones of cluster ellipticals and their supermassive black holes \citep{lbr00}.

The baryonic mass of $z<1$ galaxy clusters is dominated by primordial gas that is virialized and heated
to $10^7$ K in the potential wells of dark matter halos \citep{a10,gpf09}. A reduction of the gas  
fraction relative to the cosmic baryon fraction can be explained by a combination of star formation, supernova 
feedback, and AGN feedback \citep{bov09}. There is evidence that $z = 2$ protoclusters are X-ray deficient 
($L_X < 10^{44}$ erg s$^{-1}$) compared to clusters at lower redshift \citep{ohc05}, supporting the idea that 
they are in the process of formation and not yet virialized. 

Chandra X-ray imaging of the Spiderweb radio galaxy (PKS 1138-26, $z = 2.156$) shows extended X-ray 
emission aligned with the radio lobes \citep{chp02}, which indicates that the X-ray 
emission may come from gas that is shock-heated by the radio jet rather than virialized in 
the cluster potential. Large polarization rotation measures are found for PKS 1138-26 and 
other high-redshift radio galaxy lobes, indicating the presence of large column densities 
of ionized, magnetized gas \citep{crv97,akm98}. This gas may have its origin in radio jet-driven 
outflows \citep{nle06}.

Large quantities of shock-heated warm molecular gas are found by {\it Spitzer} in 30\% 
of 3C radio galaxies at $z<0.2$ \citep{obg10,oaa07,goe12}. The heretofore most luminous and distant H$_2$  
emission is seen in the $z=0.29$ Zw 3146 brightest cluster galaxy, with a single H$_2$ line luminosity 
$L$(H$_2$ 0-0 S(3))$= 6.1\times 10^{42}$ erg s$^{-1}$ \citep{erf06}. In radio galaxies, the H$_2$ is most 
likely powered by kinetic energy from the relativistic jet that is dissipated in the host galaxy interstellar 
medium (ISM). The large ratio of H$_2$ to PAH luminosity observed in these and other H$_2$ emission galaxies 
(MOHEGs) can not be generated in stellar photodissociation regions (PDRs). X-ray heating by relatively 
low-luminosity AGNs is also insufficient to power the observed H$_2$ rotational line luminosities \citep{obg10}. 
Star formation appears to be suppressed in radio-loud MOHEGs \citep{nbs10}, falling a factor of $\sim 50$ below the 
Kennicutt-Schmidt relation between specific star-formation rate and gas surface density \citep{k98}. 

A spectacular protocluster surrounds the Spiderweb radio galaxy \citep{krp00}. As many as 121 candidate 
cluster members within a $4.5\arcmin$ (2.2 Mpc) radius have been identified as possible Ly$\alpha$ 
emitters (37 LAE), H$\alpha$ emitters (40 HAE), or extremely red objects (44 EROs) via filter 
photometry \citep{krp00,kpr04a}. A large overdensity of sources with red IRAC [3.6]-[4.5] 
colors or bright 24 $\mu$m emission are found within a radius of $\sim 1-2\arcmin$ \citep{gsd12, mvd12}.
At least 24 cluster members have been spectroscopically detected in 
Ly$\alpha$ (15) or H$\alpha$ emission (9) and have redshifts close to that of the Spiderweb 
\citep{pkr00,kpo04b,vrm07,dtd10}. The cluster consists of two merging sub-clumps, with velocity 
dispersions of 520 and 280 km s$^{-1}$ for the associated LAEs and a combined virial mass of 
$\sim 1\times 10^{14} M_\odot$ \citep{pkr00}. Several HAEs are bound to the Spiderweb Galaxy with a velocity 
dispersion of 360 km s$^{-1}$, indicating a virial mass of $8\times 10^{13} M_\odot$\citep{kpo04b}.

Rest UV to optical continuum imaging of the cluster core with HST shows multiple satellite galaxies 
with peculiar morphologies within 150 kpc of the Spiderweb Galaxy \citep{prm98,moz06,hok09}.
An enormous Ly$\alpha$ halo imaged by the VLT surrounds the galaxy out to a radius of $> 40$ kpc, 
indicating a large reservoir of atomic gas \citep{krm02, hor08}.

We made a deep {\it Spitzer} infrared spectroscopic map of the central region of the PKS 1138-26 
protocluster, centered on the Spiderweb radio galaxy, in order to explore mid-IR indicators of star 
formation and shocked gas. Here we present mid-IR spectra of the Spiderweb and two other protocluster
galaxies which are bright in the MIPS 24 $\mu$m band. Spectra of the remaining sources will be presented 
in a future paper. Distance-dependent quantities in this paper are calculated using cosmological 
parameters $H_0=70$ km s$^{-1}$, $\Omega_\Lambda=0.7$, and $\Omega_m=0.3$.

\section{Observations}

We used the {\it Spitzer} Infrared Spectrograph \citep{h04} to map a $3.2\arcmin \times 1.1\arcmin$  
strip of the PKS 1138-26 protocluster (Fig. 1). The LL1 (20.5-38.5 $\mu$m) low-resolution slit 
($168\arcsec \times 10\farcs 7$) was stepped perpendicular to the slit in $11 \times 5 \farcs 0$ 
steps and parallel to the slit in $5 \times 5\farcs 0$ arcsecond steps to create a map centered on the 
Spiderweb radio galaxy.  Exposures were 2$\times$120 s per pointing.  The map was repeated 8 
times over 5 days, giving a total exposure time of $29.3$ hr and a mean exposure time per pixel of 4.2 hrs. 
Rotation of the spacecraft and spectrograph slit (PA$=124.6-126.3\arcdeg$) over this time period improved the 
spatial sampling of the map. Parallel data were also recorded in the LL2 (14-21 $\mu$m) slit, which will be 
reported elsewhere.

Spectra (Fig. 2) were reduced from SSC Pipeline 18.18 basic calibrated data (BCDs) using pipeline uncertainties and custom 
background subtraction. The background was estimated by stacking the BCD frames in a time series, 
masking sources, and fitting each pixel time series with a robust, low-order polynomial. The LL1 spectral cube was 
built using CUBISM \citep{sad07a}, using a spatial oversampling factor of 3, and cleaned using automatic $10\sigma$ and 
interactive bad-pixel identification and removal. A noise cube was constructed from the difference of the first and second
exposure for each pointing, and its pixel-to-pixel standard deviation was found to agree closely with the pipeline-derived 
uncertainties as a function of wavelength. For most purposes, we confidently use the pipeline uncertainties in our data analysis. 
However, these uncertainties appear to significantly underestimate the noise at observed wavelengths $>33$ $\mu$m 
(rest frame 10.4 $\mu$m). We attribute this to systematically deviant pixels that we were unable to identify and  
remove in this noisier wavelength range. Feature fluxes measured in this wavelength range, such as those of [S {\sc iv}] and the 11.3 
$\mu$m PAH, should be interpreted with caution.

Sources were identified and their positions measured in the collapsed IRS cube, weighted by the MIPS 
24 $\mu$m band spectral response (Fig. 1). Spectral extractions were performed using a 3D optimal extraction 
algorithm that we adapted from SPICE 2D optimal extraction \citep{nol07,h86}. We used a wavelength-smoothed 
and interpolated 3D PSF template derived from the Spiderweb itself for optimal extraction. The flux from this
source is dominated by the spatially unresolved AGN. Spectra were optimally
extracted (using S/N weighting) within an aperture that expands linearly with wavelength and has a
diameter of 8 pixels (2.7 native LL1 pixels, or $14\arcsec$) at 27$\mu$m. Such a small aperture is utilized to minimize
contamination from nearby sources. A wavelength-independent flux calibration factor equal to the pixel
solid angle ($5.624\times 10^{-11}$ sr/pixel) was derived from the ratio of  SPICE over CUBISM extractions 
of the Spiderweb, in BCDs where it was centered on the slit. We applied a correction for  detector fringeing 
to the Spiderweb spectrum, reducing $\sim 3\%$ oscillations in the observed 21-25 $\mu$m region down to 
the 1\% level.


We used PAHFIT \citep{sd07b} to obtain a least-squares fit to the continuum, emission lines, and PAH features. 
The continuum was fit with the sum of blackbody spectra covering a range of dust temperatures.  No extinction was 
included in the fits since it did not improve them. The emission lines were fit with unresolved Gaussian line 
profiles. In the case of emission line blends, the fluxes were summed to give the total flux of the blend, rather 
than report individual line fluxes. The PAH features are fit by Drude profiles, with rest wavelengths and widths 
fixed to values appropriate for star-forming low-redshift galaxies, but with relative intensities allowed to vary. 
While it is not clear a priori that the mean PAH composition and feature shapes should be the same at high redshift, 
this provides a good fit for most features, except for the 8.6 $\mu$m PAH which appears to be shifted to shorter
wavelength (\S 3). 

\section{Results}

We present the stacked IRS cube in Figure 1, weighted by the MIPS 24 $\mu$m band spectral response. 
We detect 13 sources with peak flux $>0.2$ mJy that yield high-quality spectra with peak S/N$>10$ per 0.18 $\mu$m bin 
in the observed 21-35 $\mu$m band. In this paper, we concentrate on the 3 brightest 24 $\mu$m sources matched to 
spectroscopically confirmed cluster members--Spiderweb, PKS 1138-26:[KPR2004] HAE 0229, and 
PKS 1138-26:[KPR2004] HAE 0131 (hereafter HAE 229 and HAE 131). 

The IRS spectrum of the Spiderweb is dominated by hot dust emission from the central AGN
(Fig. 2a,b). A silicate emission bump observed at 9.3-11$\mu$m (rest) is fit by a 120 K blackbody multiplied
by the silicate emissivity function. We detect PAH emission at 7.7 $\mu$m, 8.6 $\mu$m, and 11.2 $\mu$m. 
The 11.33 $\mu$m PAH feature falls in a noisy region of the spectrum, so it is formally not detected.
However, the majority of the reported 11.23 $\mu$m flux (Table 1) is likely contributed by the blue wing of
the 11.33 $\mu$m PAH feature. 

We detect a number unresolved emission lines (Table 2) in the spectrum of the Spiderweb, including [Ne {\sc vi}], 
[Ar {\sc iii}], [S {\sc iv}], and H$_2$ 0-0 S(3). The H$_2$ 0-0 S(4) line may be detected, but could be confused with
substructure near the peak of the 7.7 $\mu$m PAH feature. A blend of [Ar {\sc ii}] and H$_2$ 0-0 S(5) is detected at 
6.9 $\mu$m. We confirm the H$_2$ 0-0 S(3) line detection with a continuum-subtracted map of the line at $z=2.156$ 
(Fig. 1, bottom). This line is only detected at the location of the Spiderweb Galaxy, with a peak significance of 
$>6\sigma$ in the map and an integrated line detection of $14\sigma$ in the spectrum.

HAE 229 and HAE 131 both have PAH-rich spectra (Figs. 2c,d, and Table 1). We detect PAH blends at 
7.7 $\mu$m, 8.6 $\mu$m, and 11.3 $\mu$m (rest) in both galaxies. The [Ar {\sc ii}] line is detected in
HAE 229 and marginally detected in HAE 131. The [Ne {\sc vi}] and H$_2$ 0-0 S(4) lines may be detected in HAE 131, 
however it is difficult to clearly distinguish these lines from possible substructure in the 7.7 $\mu$m PAH feature.
While the H$_2$ 0-0 S(3) line is formally detected in HAE 131 ($0.11 \pm 0.03 \times 10^{-14}$ erg s$^{-1}$ 
cm$^{-2}$), the peak of the line is not at the expected wavelength and may be contaminated by a noise spike, so we treat 
this as an upper limit. This line is clearly not detected in the continuum-subtracted H$_2$ S(3) map (Fig. 1). Hot dust continuum underlies the PAH spectrum in HAE 229 and weaker continuum emission is also 
visible at the short wavelength end of the HAE 131 spectrum.

Shifting the wavelength of the 8.6 $\mu$m PAH feature from the standard wavelength of 8.61 $\mu$m to 8.57 $\mu$m 
(corresponding to a Doppler shift of -1400 km s$^{-1}$) gave a better fit to the spectra of all three sources. We believe 
this indicates a shift in the intrinsic wavelength rather than a Doppler shift because the emission lines and 7.7 and 11.3 $\mu$m 
PAH features have wavelengths consistent with the optical-NIR emission line redshifts. The 8.6/7.7 micron PAH ratios (0.21-0.38) 
seen in the three galaxies are somewhat greater than the range of PAH ratios (0.11-0.21) seen in low-redshift 
star-forming galaxies \citep{sd07b}. These differences in 8.6 $\mu$m PAH wavelength and PAH ratios may indicate a 
difference in chemical composition or excitation of the PAH molecules compared to low-redshift galaxies.

\section{Discussion}

\subsection{Spiderweb Galaxy}

The Spiderweb Galaxy is surrounded by a large number of smaller satellite galaxies, seen in
{\it HST} ACS images \citep{prm98,moz06,hok09}. Eleven cluster members or candidates have been identified within our 
$14\arcsec$ diameter spectral extraction aperture, according to NED (Fig. 1). These include the Spiderweb, X-ray source 
CXOGBA J114047.9-262906, 4 LAE candidates, 3 HAE candidates, and 2 EROs. Except for the Spiderweb, none have 
spectroscopic redshifts available. From the strong hot dust emission and low equivalent widths of the PAH features in the 
Spiderweb spectrum, it is clear that most of the observed 24 $\mu$m emission comes from the AGN. The strong
[Ne {\sc vi}] and  [S {\sc iv}] emission lines are additional indicators of a dominant AGN contribution to the spectrum. 

The Spiderweb has a 7.7$\mu$m PAH luminosity of $1.7\times 10^{45}$ erg s$^{-1}$ 
($4.4\times 10^{11} L_\odot$, Table 3).  In comparison, the brightest ULIRG in the \cite{edh11} sample has a PAH-dominated 
8 $\mu$m luminosity of $2\times 10^{11} L_\odot$, star-forming IR (SFIR) luminosity of 
$L_{SFIR}\sim 3.3\times 10^{12} L_\odot$, and a star formation rate of SFR $=500 M_\odot$ yr$^{-1}$. The 
corresponding conversions among 8 $\mu$m luminosity, total star-forming IR luminosity, and star formation rate 
are $L_{SFIR}/L_{8}=17$ and $L_{8}$/SFR$=4\times 10^{8} L_\odot M_\odot^{-1}$ yr. Lacking rest-frame far-IR fluxes, we 
adopt these as representative conversion factors for the ULIRGs in our sample. This gives an approximate PAH-based 
star-forming FIR luminosity for the Spiderweb of $L_{SFIR}\sim 7.4\times 10^{12} L_\odot$ and a star-formation rate of 
$\sim 1100 M_\odot$ yr$^{-1}$. Equally large PAH luminosities and PAH-based star-formation rates have been reported 
in the $z=3.53$ radio galaxy 6C J1908+7220 \citep[][$4000 M_\odot$ yr$^{-1}$]{sod08} the Cloverleaf 
Quasar \citep[][$\sim 1000 M_\odot$ yr$^{-1}$]{lst07}, and other high-redshift mm-bright QSOs \citep{lst08}. 

We estimate a similar star formation rate of  $\sim 1200 M_\odot$ yr$^{-1}$ from the extended 
H$\alpha$ luminosity of the Spiderweb \citep{nle06}. Here we use 
SFR(H$\alpha$, $M_\odot$ yr$^{-1}$)$=3.1 \times 10^{-8} L$(H$\alpha$) \citep{ktc94} and assume no extinction. 
The extended H$\alpha$ luminosity was measured after subtracting the point-like AGN contribution \citep{nle06}, 
but the AGN might still contribute to the extended H$\alpha$ emission via photoionization or radio-jet driven shocks. 

A much lower star formation rate of $57 \pm 8$ $M_\odot$ yr$^{-1}$ is estimated from the diffuse 32-kpc radius UV halo 
\citep{hor08}. This rate increases to 142 $M_\odot$ yr$^{-1}$ correcting for screen extinction (E(B-V)$\sim 0.1$ from 
rest UV color), and 325 $M_\odot$ yr$^{-1}$ if all of the diffuse and compact UV sources are included in the estimate 
\citep{hor08}. This is still far below the PAH-based star-formation rate, indicating that the assumption of a uniform 
dust screen is too simplistic. Instead, much of the star formation in the Spiderweb must be highly obscured ($>2.2$ mag) 
at rest-UV wavelengths. 

The Spiderweb is the brightest and most massive galaxy in the PKS 1138-26 cluster, with a stellar mass of 
$1.1\times 10^{12} M_\odot$ estimated from the SED of the central radio galaxy \citep{hok09,ssb07}. 
Adding the masses of the satellite galaxies estimated by \cite{hok09} increases the total stellar mass 
to $1.2\times 10^{12} M_\odot$ within our $7 \arcsec$ (60 kpc) radius spectral extraction aperture. 
The stellar mass doubling time scale at the PAH-inferred star formation rate is therefore $\sim 1.1$ Gyr. It 
appears that the Spiderweb is in the process of building the bulk of its stellar mass at the observed epoch.   
The Spiderweb may gain additional stellar mass (a factor of 1.1-2.2) by accreting known satellite galaxies 
within a 150 kpc radius \citep{hok09}.

We measure an H$_2$ 0-0 S(3) luminosity of $1.4 \times 10^{44}$ erg s$^{-1}$ ($3.7\times 10^{10} L_\odot$). 
We note that the H$_2$ 0-0 S(2), S(1) and S(0) lines may be similarly luminous,  but are redshifted beyond the
longest wavelength detectable with the IRS. Taken at face value, the H$_2$ 0-0 S(3) and S(5) line fluxes fit by 
PAHFIT would indicate a temperature of $\sim 920$ K and an H$_2$ mass of $\sim 1\times 10^7 M_\odot$, assuming a 
single-temperature gas in equilibrium. However, the H$_2$ 0-0 S(5) line is blended with [Ar {\sc ii}]. We estimate 
a more conservative upper limit on the H$_2$ temperature and lower limit on the H$_2$ mass by assuming that the 
[Ar {\sc ii}] $+$ H$_2$ 0-0 S(5) emission line fit gives an upper limit on the H$_2$ 0-0 S(5) line flux. This yields 
an H$_2$ temperature (for gas contributing to the S(3) and S(5) lines) of 300 K$<T<1200$ K and a warm molecular 
hydrogen mass of $>9\times 10^{6}-1\times 10^{9} M_\odot$ (Fig. 3). 

In the same temperature range, 3C radio galaxies at $z<0.22$ \citep{obg10} have warm H$_2$ masses more than 
$8$ times smaller than the Spiderweb (Fig. 3). In the lowest temperature range probed by {\it Spitzer}, their H$_2$ 
0-0 S(0)-S(2) lines indicate warm ($T=100-300$ K) H$_2$ masses $>8\times 10^{6}-2\times 10^{10} M_\odot$. It 
is likely that the Spiderweb has a much larger total warm H$_2$ mass at $T=100-300$ K than what we have detected 
with the 0-0 S(3) and S(5) lines at $T>300$ K. If the ratio of warm to total H$_2$ mass is similar to that
of low-redshift radio galaxies, then the extremely luminous H$_2$ emission from the Spiderweb could be explained by 
an extremely large molecular gas mass of $\sim2\times10^{11} M_\odot$. This would give a reasonable H$_2$ to stellar 
mass ratio of $\sim 0.2$.  A similarly large molecular gas mass of $1.4\times10^{11} M_\odot$ has been inferred from CO (1-0)
emission from the $z=2.8$ lensed quasar SMM J041327.2+102743 \citep{rcm11}.

We measure a lower limit (from S(3) only) for the ratio of the summed first four H$_2$ 0-0 pure rotational lines  
to 7.7 $\mu$m PAH of $L$(H$_2$ 0-0 S(0)-S(3))/$L$(PAH7.7)$>0.08$, more than a factor of 2 greater than the empirical 
and theoretical ratio of $<0.04$ for UV heating in a PDR. This makes the Spiderweb Galaxy the most luminous known 
radio MOHEG \citep{oaa07,obg10,goe12}, a class of galaxies where the molecular gas is heated by dissipation of 
mechanical energy. By analogy to low-redshift radio MOHEGs, the presence of a large mass of warm molecular hydrogen 
indicates that a significant fraction of the molecular gas content of the Spiderweb has been shock-heated by the 
radio jet, in turn suppressing star formation \citep{obg10,nbs10}. \cite{nle06} estimate a radio jet mechanical 
power, based on the radio luminosity of PKS 1138-26, of $\sim 10^{46}$ erg s$^{-1}$. Only $\sim 1\%$ of this would 
be enough to power the observed H$_2$ 0-0 S(3) line. On the other hand, the absorption-corrected (intrinsic 
$N_\mathrm{H}=2.6\times 10^{22}$ cm$^{-2}$) 2-10 keV X-ray luminosity of the AGN is only 
$4 \times 10^{45}$ erg s$^{-1}$ \citep{chp02}, which is insufficient to power the H$_2$ emission. The observed ratio 
$L$(H$_2$ 0-0 S(0)-S(3))/$L_X$(2-10 keV)$>0.035$ is more than 5 times greater than the theoretical maximum ratio for 
AGN X-ray heating of $<0.007$ \citep{obg10,goe12}.

From our {\it Spitzer} spectrum, the monochromatic mid-IR luminosity of the dust-obscured Spiderweb AGN 
is $\nu L_\nu$(11 $\mu$m)$= 1.9\times 10^{46}$ erg s$^{-1}$. Using an IR/UV flux conversion factor of 
$f_\mathrm{UV}=0.36$ and bolometric IR correction factor of $f_\mathrm{bol}=1.7$ for the AGN dusty torus, both derived 
from $z<1$ radio galaxies and quasars \citep{owa06}, we estimate an AGN bolometric luminosity of 
$L_\mathrm{bol}= (f_\mathrm{bol}/f_\mathrm{UV})\nu L_\nu$(11 $\mu$m)$=9.5 \times 10^{46}$ erg s$^{-1}$,
$\sim 3$ times the total IR luminosity from star formation. This is a factor of $\sim 2$ larger than the bolometric luminosity 
estimate of \cite{nbs10}, which is based on the absorption-corrected AGN X-ray luminosity \citep{chp02}. Given the 
uncertainties in the IR and X-ray bolometric corrections, we consider the two estimates to be in reasonable agreement. 
Using our mid-IR based estimate of the hidden quasar luminosity and an assumed mass-energy conversion efficiency of 0.1, 
we estimate a mass accretion rate of $\sim 16 M_\odot$ yr$^{-1}$ for the supermassive black hole that powers the Spiderweb 
AGN. This is $\sim 1\%$ of the PAH-based instantaneous star formation rate integrated over the entire galaxy. 


\subsection{HAE 229 and HAE 131}

HAE 229 ($z=2.149$) and HAE 131 ($z=2.152$) are two of the brightest H$\alpha$ emission candidates in the 
survey of the PKS 1138-26 protocluster by \cite{kpr04a}, and have confirmed H$\alpha$ and [N {\sc ii}] 
emission in their NIR (rest-frame optical) spectra \citep{kpo04b}. The projected distances from the center 
of the Spiderweb Galaxy are $32\farcs 9$ (280 kpc) and $49\farcs 9$ (420 kpc), respectively.

Note that the QSO CXO J114045.9-262916 (PKS 1138: [KPR2004] HAE 0215, $z=2.157$) falls within the spectral 
extraction aperture $5\farcs4$ SW of HAE 229,  (Fig. 1). The positional uncertainty for the QSO is $0\farcs45$, 
and we estimate an absolute positional uncertainty of $<1\farcs0$ for HAE 229 from our IRS spectral map. The 
two sources are spatially resolved at the wavelength of the 7.7 $\mu$m PAH band. By comparing the PAH feature 
and continuum extracted at the peak pixel of each source, we estimate that $<30\%$ of the 7.7$\mu$m PAH flux and
$<60\%$ of the 8 $\mu$m continuum in the HAE 229 spectrum can be contributed by the QSO. 

We measure a PAH luminosity for HAE 229 of $3.5\times 10^{11} L_\odot$. The inferred star-forming IR luminosity is 
$L_{SFIR}\sim 6.0\times 10^{12} L_\odot$ and the PAH-derived star formation rate is $\sim 880 M_\odot$ yr$^{-1}$, 
similar to the Spiderweb Galaxy. HAE 229 is one of two red galaxies in the PKS 1138-26 cluster that were detected in 
H$\alpha$ by \cite{dtd10} (galaxy 464). The H$\alpha$ luminosity \citep{kpo04b} indicates a star formation rate
of $=20$ $M_\odot$ yr$^{-1}$, a factor of $\sim 40$ less than what we estimate from the 7.7 $\mu$m PAH feature (Table 3).
This indicates that the bulk of star formation activity is heavily obscured ($>4.1$ mag) at the 
wavelength of H$\alpha$. SED fitting yields a stellar mass of $5\times10^{11} M_\odot$, and an age of 
2.4 Gyr \citep{dtd10}. The time required to double the stellar mass is only 600 Myr, indicating that 
protocluster galaxy HAE 229 is undergoing a major event in its construction.

We measure a PAH luminosity for HAE 131 of $1.9\times 10^{11} L_\odot$, half as bright as HAE 229 and
the Spiderweb. The corresponding estimated IR luminosity and star formation rate are 
$L_{SFIR}\sim 3.2\times 10^{12} L_\odot$ and $\sim 470 M_\odot$ yr$^{-1}$. The H$\alpha$ luminosity \citep{kpo04b} 
indicates a star formation rate of  34 $M_\odot$ yr$^{-1}$, a factor of 14 less than what we estimate from 
the 7.7 $\mu$m PAH feature (Table 3). As for HAE 229, most of the star formation in HAE 131 must be obscured 
($>2.9$ mag) at the wavelength of H$\alpha$. 

The ultraluminous PAH emission and large PAH/H$\alpha$ flux ratios in both HAE 229 and HAE 131 clearly indicate that 
they are ULIRGs, undergoing a rapid burst of obscured star formation. Similar PAH luminosities and equivalent widths
have been seen in other high-redshift, 24 $\mu$m-selected ULIRGs \citep{sya07}. \cite{tdk11} find that HAEs in the PKS 1138-26 
cluster have systematically smaller H$\alpha$/K-band stellar continuum ratios than HAEs in the 4C 23.56 cluster 
($z=2.483$), attributing this to smaller specific star formation rates. However, we find H$\alpha$ extinction factors 
in two PKS 1138-26 HAEs of 14 and 40, compared to a range of 2-14 in the 4C 23.56 cluster, based on H$\alpha$/24 $\mu$m 
fluxes. If this tendency towards larger extinction in PKS 1138-26 HAEs holds up, the difference in H$\alpha$/K-band 
stellar continuum ratios may be ascribed to differences in extinction, rather than in specific star formation rates.

The lack of high-ionization lines in the HAE 229 spectrum indicates a relatively weak AGN contribution, though there is 
an indication of hot dust emission at at 6-7 $\mu$m rest. Detection of [Ne {\sc vi}] in the HAE 131 spectrum 
together with the detection of hot dust  indicates an AGN contribution, though it must be relatively weak, considering 
the large PAH equivalent widths. We do not detect H$_2$ 0-0 S(3) emission in HAE 229 nor HAE 131, with a 2$\sigma$ 
upper limit of $< 1\times 10^{-15}$ erg s$^{-1}$ cm$^{-2}$ ($<9\times 10^{9} L_\odot$), even though they have similar PAH 
luminosities and inferred star formation rates to the Spiderweb. This supports the idea that the radio jet or AGN 
excites the H$_2$ emission in the Spiderweb, rather than star formation.

\section{Conclusions}

We have found PAH emission indicating star formation rates of $500-1100 M_\odot$ yr$^{-1}$ in three galaxies
in the PKS 1138-26 radio galaxy protocluster, including the Spiderweb central radio galaxy, HAE 229, and 
HAE 131. The corresponding stellar mass doubling time scales in the Spiderweb and HAE 229 are 1.1 Gyr and 
$600$ Myr, respectively. This indicates that these protocluster galaxies are undergoing a period of rapid 
growth and that the PKS 1138-26 protocluster is a massive galaxy construction zone. Similarly high
star formation rates have been observed in other high-redshift radio galaxies \citep[e.g.][]{sod08}, in
contrast to low-redshift ($z<0.2$) radio galaxies where star formation rates are generally lower 
($<5 M_\odot$ yr$^{-1}$), in spite of the presence of up to $10^{10} M_\odot$ of molecular gas in the
central few kpc \citep{sor07, obg10,ssp11}. Low-redshift radio jets appear to be more effective at 
suppressing star-formation in their host galaxies. However, the dynamical state and gas physical 
properties of low-redshift radio galaxies are also much more evolved, as indicated by the 
presence of virialized hot gas in their halos and surrounding galaxy clusters.

Star formation rates estimated from the 7.7 $\mu$m PAH feature are a factor of 14-40 greater than
estimates based on H$\alpha$, for HAE 229 and HAE 131. This indicates that star formation is heavily
obscured (by $>2.9-4.1$ mag) at H$\alpha$ and shorter wavelengths. This is in contrast to the Spiderweb,
where the PAH-based and H$\alpha$-based star formation rate estimates are comparable. This may suggest
a different mode of more-extended, less obscured star formation in the Spiderweb. This in turn may be
consistent with the observed extended UV continuum emission \citep{hor08} if the UV extinction
is more complicated than a simple obscuring screen. 

The extended mode of star-formation in the Spiderweb could possibly be a result of negative feedback 
quenching star-formation in the nucleus, coupled with accretion of many satellite galaxies enhancing
star formation in the outer regions of the galaxy. High luminosity H$_2$ 0-0 S(3) emission from the Spiderweb 
indicates a large mass of warm ($T> 300$ K) molecular hydrogen, most likely heated by radio jet feedback from the 
AGN, which may have already quenched nuclear star formation and may eventually quench quasar activity. 
In other words, the high-star formation rates observed in high-redshift radio galaxies may be attributed
to the accretion of gas and gas-rich satellite galaxies rather than nuclear star formation. At low redshift,
the gas has been virialized and heated to $10^{7}$ K and the satellites have been incorporated into the 
central radio galaxy, such that there is no longer a steady gas supply for star formation. The remaining gas
that is able to cool and make its way into the central kpc-scale disk is unable to form stars efficiently
because it is either continuously or episodically heated by the radio jet. 

Herschel FIR-sub-mm photometry will determine the total IR luminosities of the galaxies in the PKS 1138-26 protocluster, 
yielding more accurate star formation rates and dust mass estimates. Herschel PACS spectroscopy may be sensitive
enough to detect the H$_2$ S(0) line and thereby measure the total mass of warm molecular gas in the Spiderweb.
The extreme luminosity of H$_2$ rotational line emission in the Spiderweb provides a strong incentive for 
launching a cooled, large aperture FIR space telescope to detect these lines and use them to measure the impact
of kinetic energy dissipation on molecular gas and star formation in even higher-redshift galaxies
\citep{aab09}. 

\acknowledgements

This work is based on observations made with the {\it Spitzer} Space Telescope, which is operated by 
the Jet Propulsion Laboratory, California Institute of Technology under NASA contract 1407.
This work benefited greatly from the NASA/IPAC Extragalactic Database (NED), which is operated by
the Jet Propulsion Laboratory, California Institute of Technology, under contract with NASA.
N. S. is a recipient of an Australian Research Council Future Fellowship.

\newpage

\begin{deluxetable}{cccccccc}
\tabletypesize{\scriptsize}
\tablecaption{PAH Features}
\tablewidth{0pt}
\tablehead{
\colhead{Source} &  \colhead{ 7.42\tablenotemark{a}} &  \colhead{ 7.60} &  \colhead{ 7.85} &  \colhead{ 8.33} &  \colhead{ 8.57\tablenotemark{b}} &  \colhead{11.23} &  \colhead{11.33}}
\startdata
Spiderweb& 2.30 (0.28) & \nodata     & 2.62 (0.07) &  \nodata     & 1.15 (0.04) & 1.08 (0.06) & 0.29 (0.14) \\
HAE 229  & 1.13 (0.57) & 0.94 (0.08) & 1.88 (0.07) & 0.28 (0.06)  & 0.62 (0.04) & $<0.14$     & 1.82 (0.16) \\
HAE 131  & $<0.56$     & 0.95 (0.10) & 1.16 (0.07) & $<0.14$      & 0.79 (0.06) & $<0.30$     & 2.24 (0.27) \\
\enddata
\tablenotetext{a}{Flux and $1\sigma$ uncertainty or $2\sigma$ upper limit 
($10^{-14}$ erg s$^{-1}$ cm$^{-2} = 10^{-17}$ W m$^{-2}$).}
\tablenotetext{b}{PAH feature at 8.6 $\mu$m is better fit using a central wavelength of 8.57 $\mu$m than 
the standard value of 8.61 $\mu$m .}
\end{deluxetable}

\begin{deluxetable}{lcccccc}
\tabletypesize{\scriptsize}
\tablecaption{Emission Lines}
\tablewidth{0pt}
\tablehead{
 \colhead{Source} & \colhead{ H$_2$ 0-0 S(5)+[Ar {\sc ii}] } & \colhead{ [Ne {\sc vi}] } & \colhead{H$_2$ 0-0 S(4) } & \colhead{[Ar \sc{iii}] } & \colhead{H$_2$ 0-0 S(3) } & \colhead{  [S {\sc iv}] }  }
\startdata
Spiderweb& 0.80 (0.08) & 1.56 (0.04) & 0.61 (0.04) & 0.37 (0.03) & 0.41(0.03) & 1.55 (0.04) \\
HAE 229  & 0.44 (0.07) & $<0.30$     & $<0.21$     & \nodata     & $<0.06$    & $<0.17$     \\
HAE 131  & 0.17 (0.08) & 0.18 (0.04) & 0.20 (0.04) & $<0.07$     & $<0.11$    & $<0.08$     \\
\enddata
\tablenotetext{a}{Notes. Flux and $1\sigma$ uncertainty or $2\sigma$ upper limit ($10^{-14}$ erg s$^{-1}$ cm$^{-2} = 10^{-17}$ W m$^{-2}$).}
\end{deluxetable}

\begin{deluxetable}{cccccc}
\tabletypesize{\scriptsize}
\tablecaption{Luminosities and SFRs}
\tablewidth{0pt}
\tablehead{
 \colhead{Source} &  \colhead{ $L$(PAH7.7)\tablenotemark{a}} &  \colhead{$L$(H$\alpha$)\tablenotemark{b}} & \colhead{$L_{SFIR}$\tablenotemark{c}} &  \colhead{SFR(PAH)\tablenotemark{d}} & \colhead{SFR(H$\alpha$)\tablenotemark{e}}}
\startdata
Spiderweb& 4.4 (0.3) & 38 (3)      & 7.4 (0.5) & 1100  (60) & 1200 (100)\\ 
HAE 229  & 3.5 (0.5) & 0.64 (0.17) & 6.0 (0.9) &  880 (130) &   20 (5)  \\ 
HAE 131  & 1.9 (0.2) & 1.1 (0.2)   & 3.2 (0.3) &  470  (30) &   34 (6)  \\ 
\enddata

\tablenotetext{a}{.Luminosity ($10^{11} L_\odot$) of 7.7$\mu$m PAH blend, including contributions from the 
                  7.42 $\mu$m, 7.60 $\mu$m, and 7.85 $\mu$m features.}
\tablenotetext{b}{H$\alpha$ luminosity ($10^9 L_\odot$). H$\alpha$ fluxes for HAE 229 and HAE 131 are from \cite{kpo04b}.
                  Spiderweb extended narrow H$\alpha$ flux has AGN contribution subtracted \citep{nle06}.}
\tablenotetext{c}{Estimated Star-forming total IR luminosity $L_{SFIR}$($10^{12} L_\odot$)$=17\times L$(PAH7.7) (Elbaz et al. 2011).}
\tablenotetext{d}{PAH-based star formation rate SFR(PAH, $M_\odot$ yr$^{-1}$)$= L$(PAH7.7)/$4\times10^8$  (see text).} 
\tablenotetext{e}{SFR(H$\alpha$, $M_\odot$ yr$^{-1}$)$=3.1 \times 10^{-8} L$(H$\alpha$) \citep{ktc94}.} 
\end{deluxetable}

\begin{figure}[ht]
  \plotone{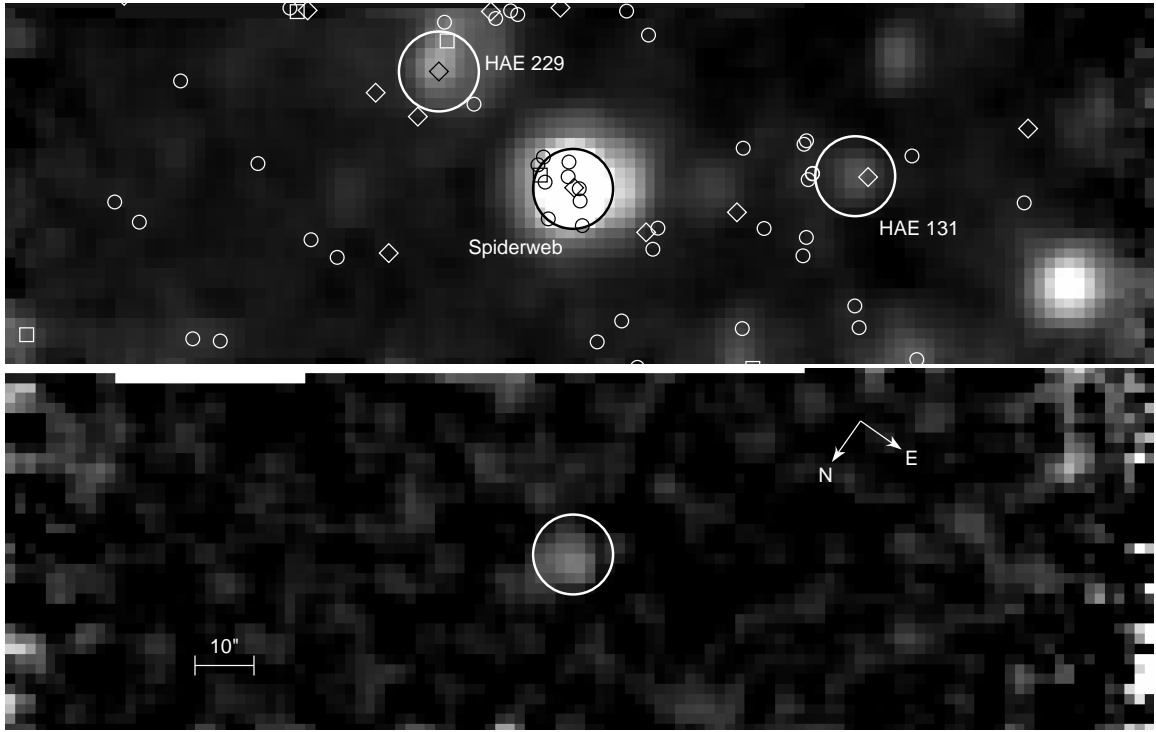}
  \figcaption{{\it Spitzer} IRS maps ($3.2\arcmin \times 1.1\arcmin$) of the PKS 1138-26 protocluster. The mean slit
    orientation and long axis of each map is at PA$=125.1\arcdeg$. Top: IRS map in synthetic MIPS 24 $\mu$m band.  The 
    locations of spectroscopic cluster members (diamonds) or candidates (small circles) and X-ray sources (squares) are 
    indicated. Spectral extraction apertures (large circles) are shown for the Spiderweb (center), HAE 229 (top), and 
    HAE 131 (right). Note that the QSO CXO J114045.9-262916 ($z=2.157$) falls inside and near the SW edge of the spectral 
    extraction aperture for HAE 229, and may contribute flux to its spectrum (\S4.2). Bottom: synthetic H$_2$ 0-0 S(3) 
    narrow-band, continuum subtracted map. The same compass rose and scale apply to both maps.}
\end{figure}

\begin{figure}[ht]
  \plotone{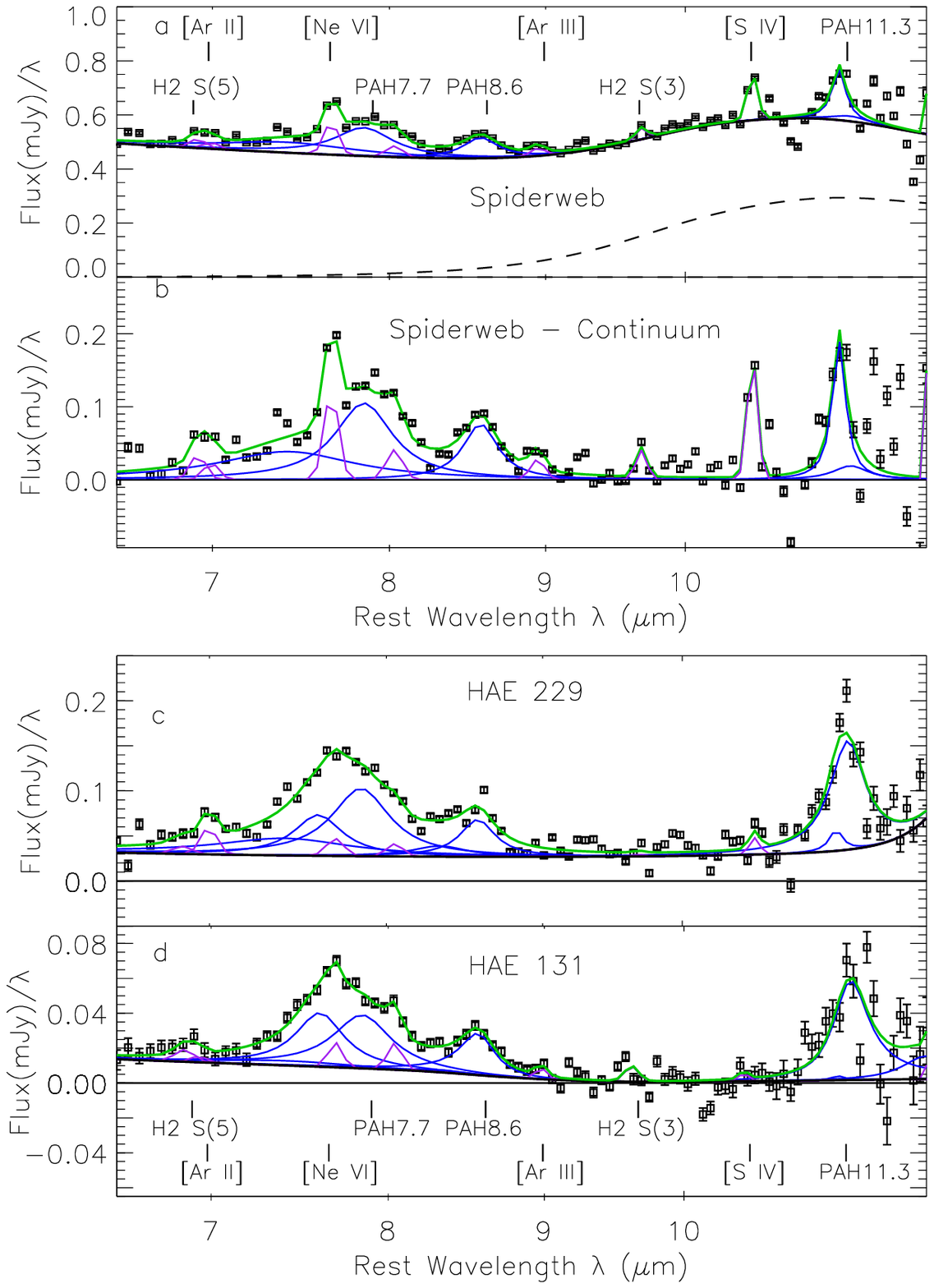}
  \vfil
  \eject
  \figcaption{See above for figure caption, which will not fit on the page.}
\end{figure}

\vfill \eject
\newpage

{\bf Figure 2 Caption}

   {\it Spitzer} IRS LL1 spectra. 
   (a) Spiderweb radio galaxy ($z=2.156$). PAHFIT spectral fit (green) and fit components: continuum (solid black),
       silicate emission (dashed), PAH features (blue), and emission lines (purple).  
   (b) Spiderweb PAH and emission line spectrum after subtracting continuum fit. 
    Emission lines [Ar {\sc ii}]+H$_2$ 0-0 S(5), [Ne {\sc vi}], [Ar {\sc iii}], H$_2$ 0-0 S(3), and [S {\sc iv}]; 
    and 7.7, 8.6, and the blue wing of the 11.3 $\mu$m PAH are detected. (c) HAE 229 ($z=2.149$) (d) HAE 131 ($z=2.152$).
    PAH features are detected at 7.7, 8.6, and 11.3 $\mu$m in both HAE 229 and HAE 131.
    The [Ar {\sc ii}] emission line is also detected.
\newpage

\begin{figure}[ht]
  \plotone{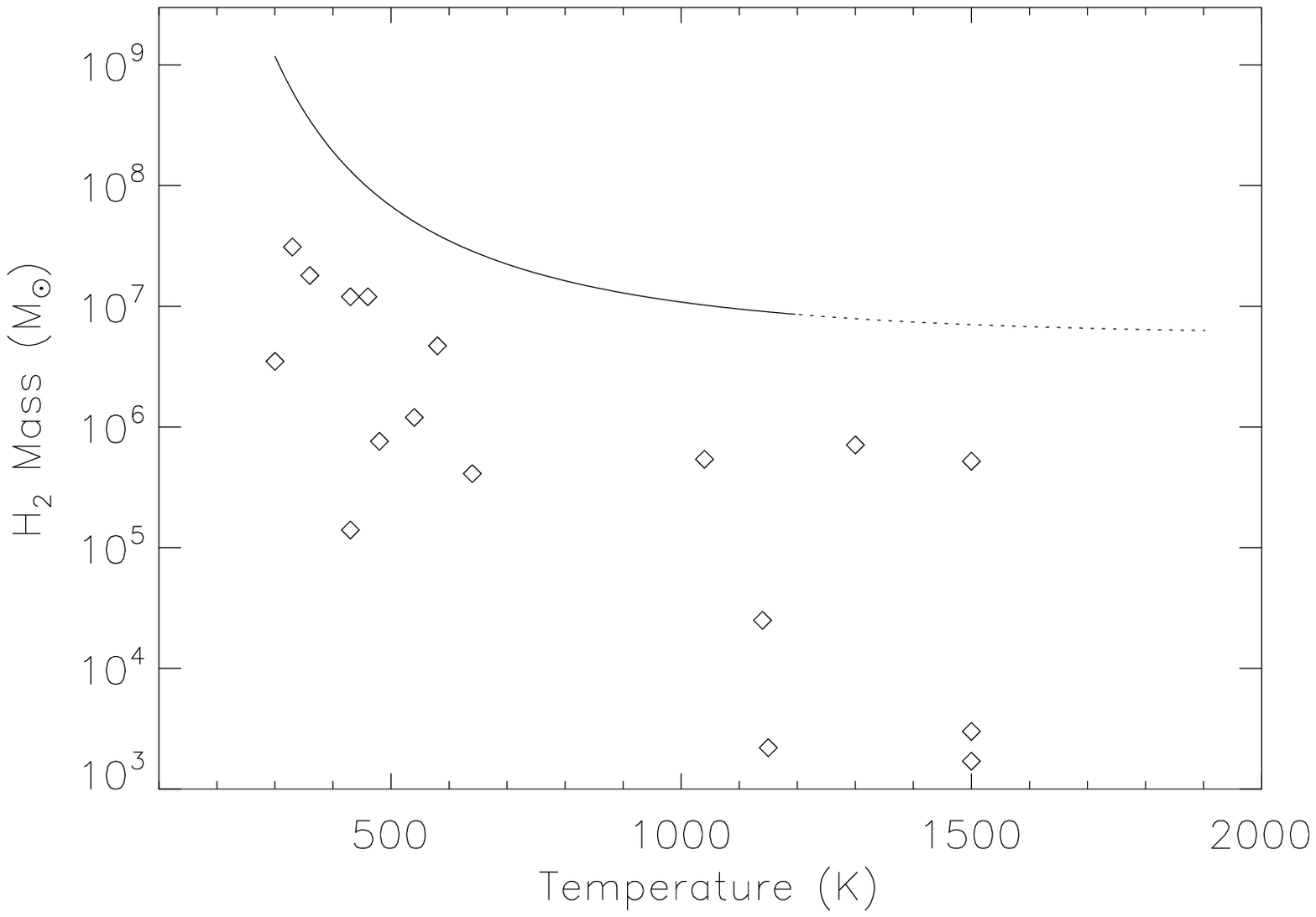}
  \figcaption{Estimated mass of warm ($300<T<1200$ K) H$_2$ in the Spiderweb Galaxy versus temperature (solid line). 
   Temperatures $T>1200$ K (dotted line) are excluded by the upper limit to the H$_2$ S(5) flux . In comparison, warm 
   molecular hydrogen masses in the same temperature range for $z<0.22$  3C radio galaxy MOHEGs \citep{obg10} 
   are also plotted (diamonds).}
\end{figure}
\end{document}